\documentclass[twocolumn]{aastex63}

\shorttitle{Spallation in X-ray bursts}
\shortauthors{Randhawa et al.}
\graphicspath{{./}{figures/}}

\begin{document}

\title{Spallation-altered accreted compositions for X-ray bursts: Impact on ignition conditions and burst ashes}

\correspondingauthor{J. S. Randhawa}
\email{randhawa@nscl.msu.edu}

\author[0000-0001-6860-3754]{J. S. Randhawa}
\affiliation{Department of Astronomy and Physics, Saint Mary’s University, 923 Robie Street, Halifax, NS, B3H 3C3, Canada}
\affiliation{National Superconducting Cyclotron Laboratory,
Michigan State University, East Lansing, MI 48824, USA}

\author{Z. Meisel}
\affiliation{Institute of Nuclear and Particle Physics, Department of Physics \& Astronomy, Ohio University, Athens, OH 45701, USA}


\author[0000-0002-9814-0719]{S. A. Giuliani}
\affiliation{National Superconducting Cyclotron Laboratory,
Michigan State University, East Lansing, MI 48824, USA}

\author{H. Schatz}
\affiliation{National Superconducting Cyclotron Laboratory,
Michigan State University, East Lansing, MI 48824, USA}
\affiliation{Department of Physics and Astronomy, Michigan State University,East Lansing, MI 48824,USA}
\affiliation{JINA Center for the Evolution of the Elements,Michigan State University,East Lansing,MI48824,USA}

\author{B. S. Meyer}
\affiliation{Department of Physics and Astronomy, Clemson University, Clemson, SC 29634-0978, USA}

\author{K. Ebinger}
\affiliation{GSI Helmholtzzentrum fur Schwerionenforschung, D-64291 Darmstadt, Germany}
\author{A. A. Hood}
\affiliation{Department of Physics and Astronomy, Louisiana State University, Baton Rouge, Louisiana 70803}
\author{R. Kanungo}
\affiliation{Department of Astronomy and Physics, Saint Mary’s University, 923 Robie Street, Halifax, NS, B3H 3C3, Canada}



\begin{abstract}
Dependable predictions of the X-ray burst ashes and light curves require a stringent constraint on the composition of the accreted material as an input parameter. Lower metallicity models are generally based on a metal deficient donor and all metals are summed up in CNO abundances or solar metal distribution is assumed. In this work, we study the alteration of accreted composition due to spallation in the atmosphere of accreting neutron stars considering a cascading destruction process. We find that the inclusion of the cascading process brings  the replenishment of CNO elements and overall survival probability is higher compared to isolated destruction of CNO elements. Spallation model provides the distribution of metals as a function of mass accretion rate. 
Multi-zone X-ray burst models calculated with reduced metallicities have enhanced abundances for high-mass nuclei in X-ray burst ashes. The increased metallicity due to the replenishment of CNO elements changes the composition of burst ashes compared to lower metallicity conditions. This will modify the thermal and compositional structure of accreted neutron star crusts.

\end{abstract}



\section{Introduction} \label{sec:intro}
Thermonuclear explosions on the surface of accreting
neutron stars in low-mass X-ray binaries give rise
to Type~I X-ray bursts \citep{Lewin93,sch06}. These are among the most frequent thermonuclear explosions in nature with the recurrence time ranging from hours to days \citep{stroh06}. The bursts are powered
by different sequences of nuclear reactions such as triple-$\alpha$ process, $\alpha p$-process and $rp$-process \citep{WW81,Cyburt2016,sch06}.  The importance of these sites as a probe for neutron star properties has been emphasized from time to time \citep{Cumming06, Zamfir12,Ozel16}. To explain the observations of X-ray bursts in terms of neutron star properties, reliable burst models are required. These models depend on a range of input parameters, where the composition of accreted material is one such major parameter \citep{Woosley04, Heger07}. The initial composition plays a vital role in deciding the ignition conditions and evolution during the burst \citep{Cumming2000, WW81, Schatz01, Woosley04, Jose10}. The choice of accreted composition in X-ray burst models varies from solar-metallicity to metal-deficient (Z$\sim10^{-3}$) \citep{Cumming2000, Schatz01, Heger07}, which reflects  the surface composition of the companion star, without any further alteration. It has been discussed previously that proton induced spallation of the accreted material in the neutron star atmospheres can potentially change the accreted composition before it settles down to the deeper layers of the neutron star \citep{Bil92}. Due to Coulomb collisions with atmospheric electrons, elements heavier than helium and hydrogen thermalize at shallower depths. At these depths incoming protons still have high enough energies to destroy these heavier elements through nuclear spallation reactions.  In \citet{Bil92} it was explicitly mentioned that the spallation of the thermalized ions by protons leads to nuclear fragments which can further undergo fragmentation, hence resulting in a cascading destruction process. Due to a lack of knowledge of the relevant spallation cross sections at that time only the isolated destruction of CNO elements was discussed. To obtain a realistic  final composition, the full cascading destruction process needs to be considered.

In this work, we calculate the  accreted composition considering spallation in the atmospheres in a full cascading destruction model. The main motive of this work is to understand the impact of replenishment in a cascading process on the survival probability of the CNO elements. The cascading process provides an accretion rate dependent metal distribution. 
\color{black}
Impact of the spallation-altered composition on the burst ignition conditions is examined. Multizone X-ray burst studies with new composition were performed resulting in a substantially modified abundance distribution for X-ray burst ashes.

\section{Method} \label{sec:method}
The kinetic energy of  free falling material onto a 1.4 $M_{\odot}$ neutron star of 10 km radius is $\sim$ 200 MeV/u. 
Due to Coulomb collisions with atmospheric electrons, heavier elements thermalize at shallower depths and their downward journey is then dictated by  diffusion. At these stopping depths, incoming protons still have high energies. The time these relatively heavy elements take to diffuse from their stopping depths to proton stopping depths is what we refer to as exposure time ($\mathrm{t_{exposure}}$) in the present study. This is the time duration for which elements are exposed to high energy protons. 

Following the discussion in \citet{Bil92}, $\mathrm{t_{exposure}}$ can be written as
\begin{equation}
\mathrm{t_{exposure}}= \frac{R}{j_{p}}\left (1-\frac{A}{Z^{2}}  \right ) y_{s}(p),
\end{equation}
where $j_{p}$ is the proton beam current, R is the factor obtained from diffusion calculations \citep{Bil92}, $A$ and $Z$ are the mass number and atomic number of a given element, respectively, and $y_{s}(p)$ is  the electron column density needed to stop  protons.
 This can be re-expressed in terms of the accretion rate as
\begin{equation}
\mathrm{t_{exposure}} \approx \frac{0.5 (kg\ cm^{-2})}{\dot{m} (kg\ cm^{-2}\ s^{-1}) } \left (1- \frac{A}{Z^{2}} \right ) .
\end{equation}
\begin{figure}
\begin{center}
\includegraphics[angle=0,scale=.68]{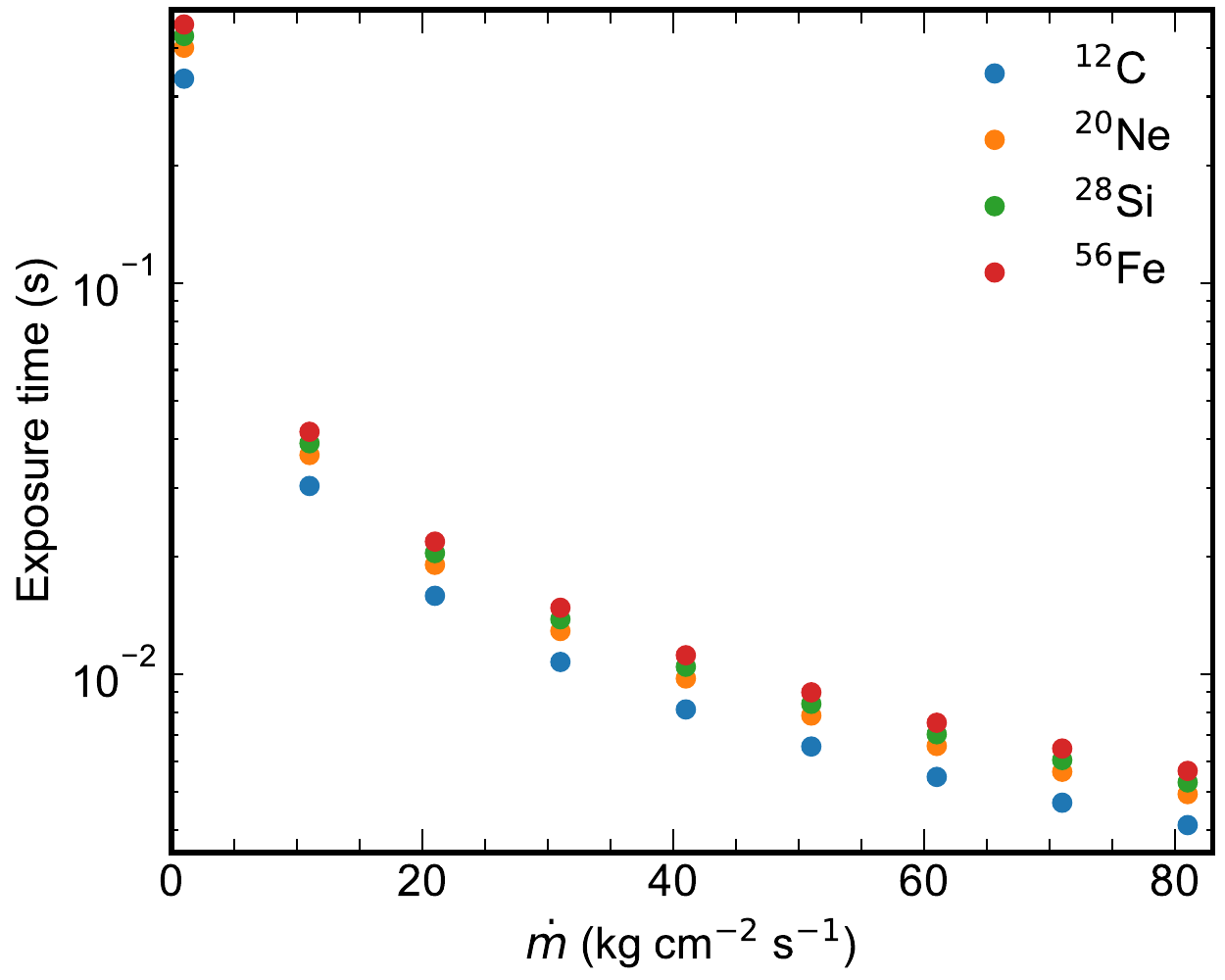}
\caption{Exposure time for different elements as a function of mass accretion rates.}
\label{figure1}
\end{center}
\end{figure}
$\mathrm{t_{exposure}}$  for C, Ne, Si and Fe elements over a range of mass accretion rates are shown in Figure~\ref{figure1}. There are two observations worth noting. First, for a given element the exposure time decreases with increasing mass accretion rate. 
Second, for a given mass accretion rate the exposure time for various elements is nearly the same. For example, in Figure~\ref{figure1} we show that the change in exposure time is very small while going from $^{12}$C to $^{56}$Fe. Therefore, for a given mass accretion rate we have assumed the same exposure time(s) for all the elements, taken as the exposure time of $^{12}$C.

The spallation process is modeled using the NucNet Tools single zone reaction network\footnote{https://sourceforge.net/p/nucnet-tools/home/Home/}. Along with the proton induced spallation reactions, $\beta$-decays were also included as the fragments produced could be highly unstable. The single-zone network includes 486 isotopes from H to Fe coupled by a total of  13076 reactions. Out of these reactions,  1421 are weak reactions and and rest are spallation reactions. 
Spallation reactions were treated as decay reactions, since the protons are constantly supplied by accretion, in order to incorporate them easily in the reaction network.
 $\beta$-decay rates were taken from the Nuclear Wallet Cards \citep{Tuli11}. The spallation reaction rates were calculated  as $j_{p}\times \sigma(E_{p})$
where $\sigma(E_{p})$ are energy dependent partial cross sections.  
The spallation cross sections $\sigma(E_{p})$ were calculated using the open source subroutines from the work of \citet{Sil98} and also includes updates of energy dependent scaling  to match the available experimental data \citep{Gallo19}. For a given mass accretion rate, abundances were evolved for corresponding exposure time.

Proton-induced spallation, as discussed above, is based on the fact that the protons are stopped in deeper layers compared to heavier elements. Hydrogen and helium have the same stopping depths (same $A/Z^{2}$), therefore, no He spallation is considered in the present case. Helium can still be spalled  while it is being slowed down in the atmosphere  but products quickly re-assembles through various reactions to restore the initial amount of helium \citep{Bil93}.

\section{Spallation results} \label{sec:floats}
\begin{figure}
 \begin{center}
\includegraphics[angle=0,scale=.55]{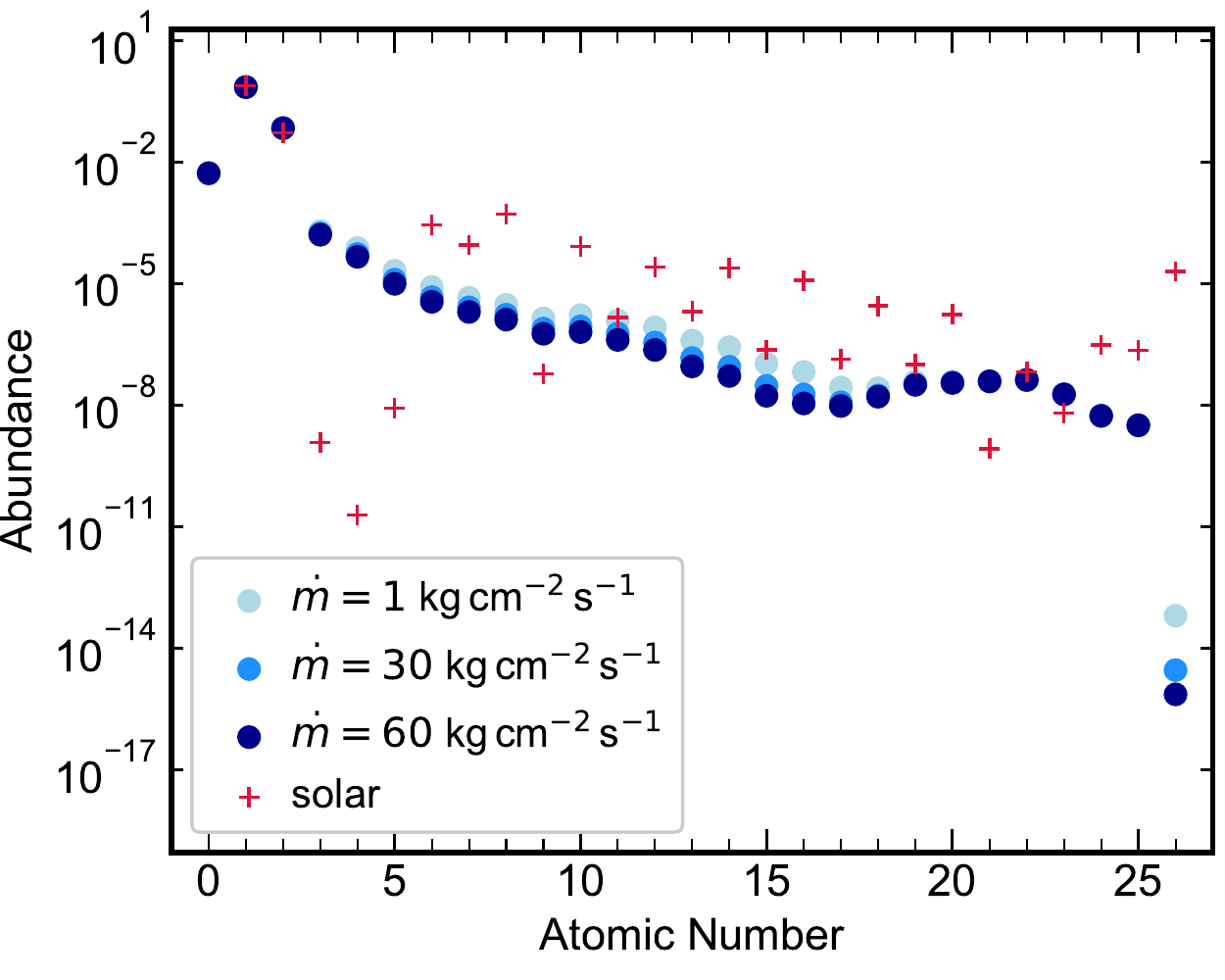}
\caption{Accreted composition with and without spallation.
Red crosses show the solar composition whereas blue dots show the final compositions surviving after material of solar composition goes through destruction process via spallation reactions, for three different mass accretion rates.}
\label{fig:SpallationProducts}
\end{center}
\end{figure}

The abundances of the elements after spallation are shown in Figure~\ref{fig:SpallationProducts} for three different exposure times corresponding to three different accretion rates. From these results we conclude  that spallation can significantly reduce the accreted abundance from solar to sub-solar for all the elements heavier than boron. 
 This change is much larger for higher mass regions. 
  


\begin{figure}

        \begin{center}
        \includegraphics[scale =0.25]{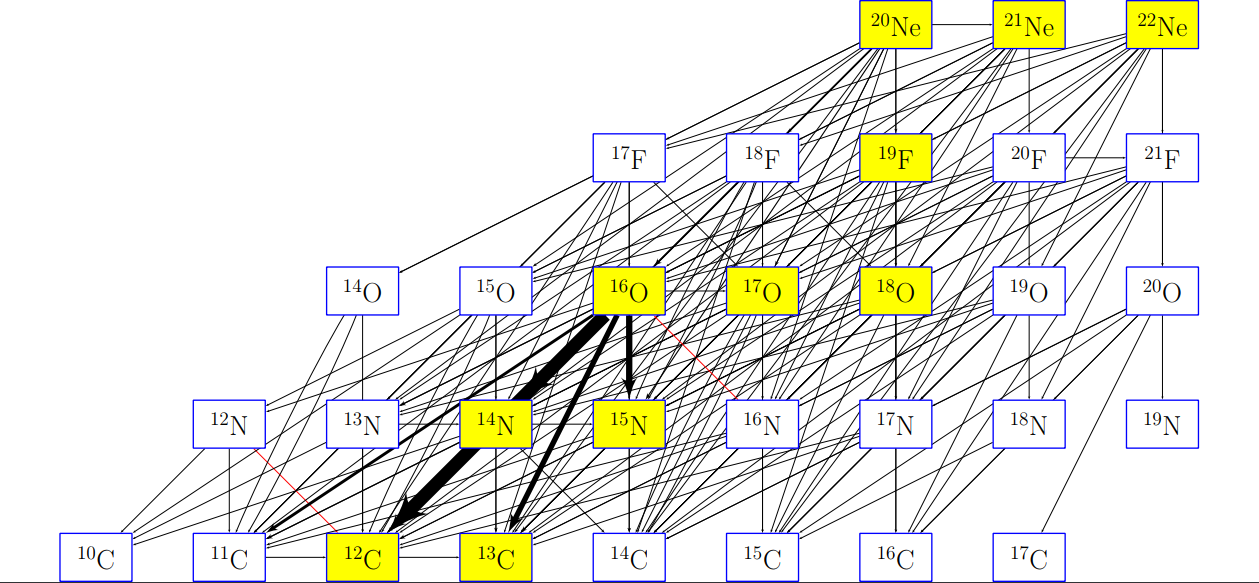}
        \caption{Time integrated net reaction flow.}
        \label{figure2.1}
        \end{center}

\end{figure}

\begin{figure}
    
        \begin{center}
        \includegraphics[scale =0.48]{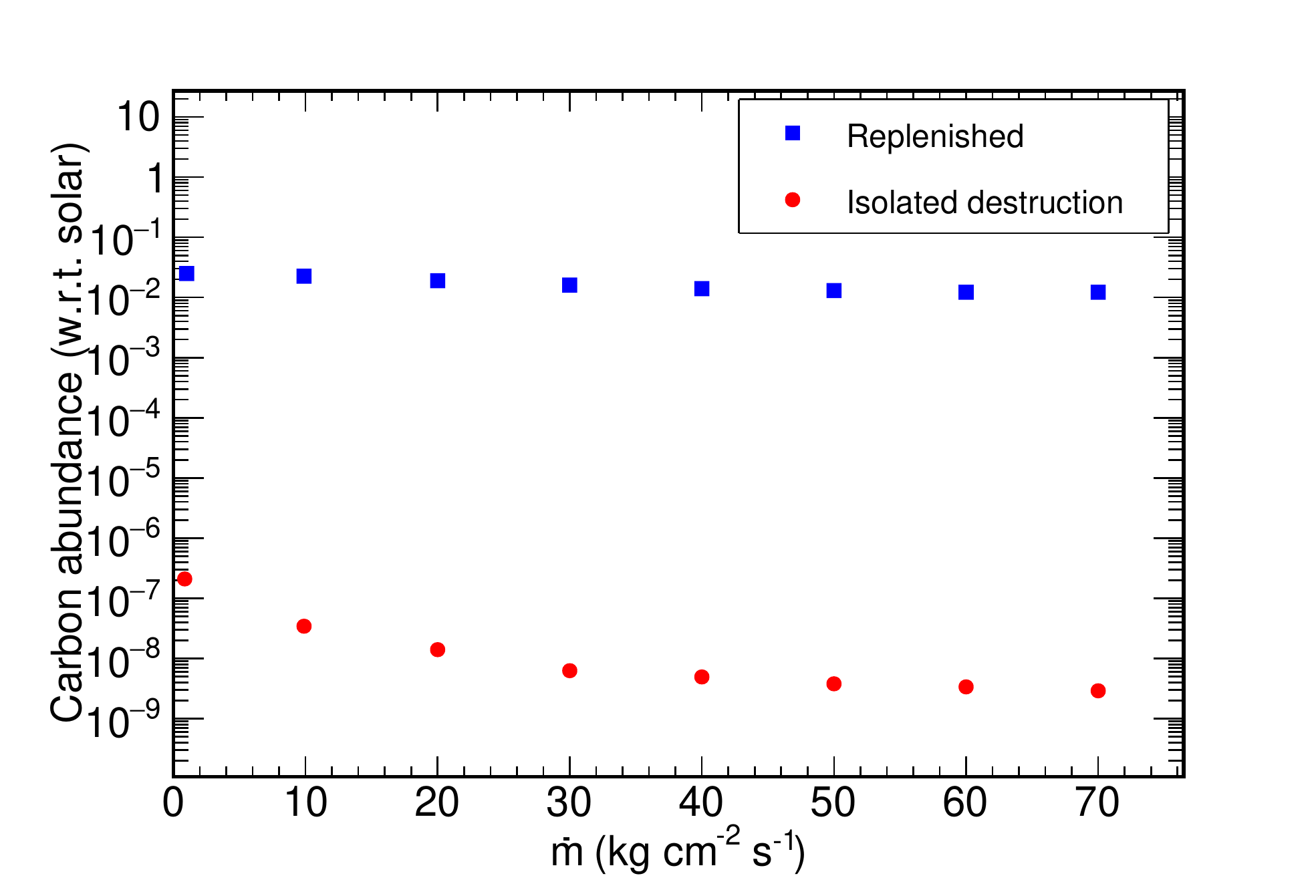}
        \caption{Abundance of carbon surviving in two different models as a function of mass accretion rates.}
        \label{figure3}
        \end{center}

\end{figure}

In \citet{Bil92}, destruction of individual CNO elements was considered (isolated destruction). However, in the cascading destruction process, due to the transformation of heavier elements into CNO elements, the CNO elements are replenished. To check whether accounting for the full cascading process makes any difference compared to isolated destruction, we followed the abundance evolution of  $^{12}$C in an isolated destruction process and in a full cascading process. 

 The time integrated net reaction flow is shown in Figure~\ref{figure2.1} for isotopes from Ne to carbon. The time integrated reaction flow shows how carbon is replenished via destruction of Ne, F, O and N. The major path of $^{12}$C production is via spallation of $^{14}$N and $^{16}$O. These production channels will be missing in the isolated destruction scenario. 

 \color{black}
  Figure~\ref{figure3} shows  that  difference in carbon abundance is conspicuous for cascading and isolated destruction models for all mass accretion rates. The final carbon abundance in the isolated destruction scenario is several orders of magnitude smaller. These results demonstrate that the CNO  destruction can be overestimated if no replenishment is considered. 

\begin{figure}
\begin{center}

\includegraphics[angle=0,scale=.48]{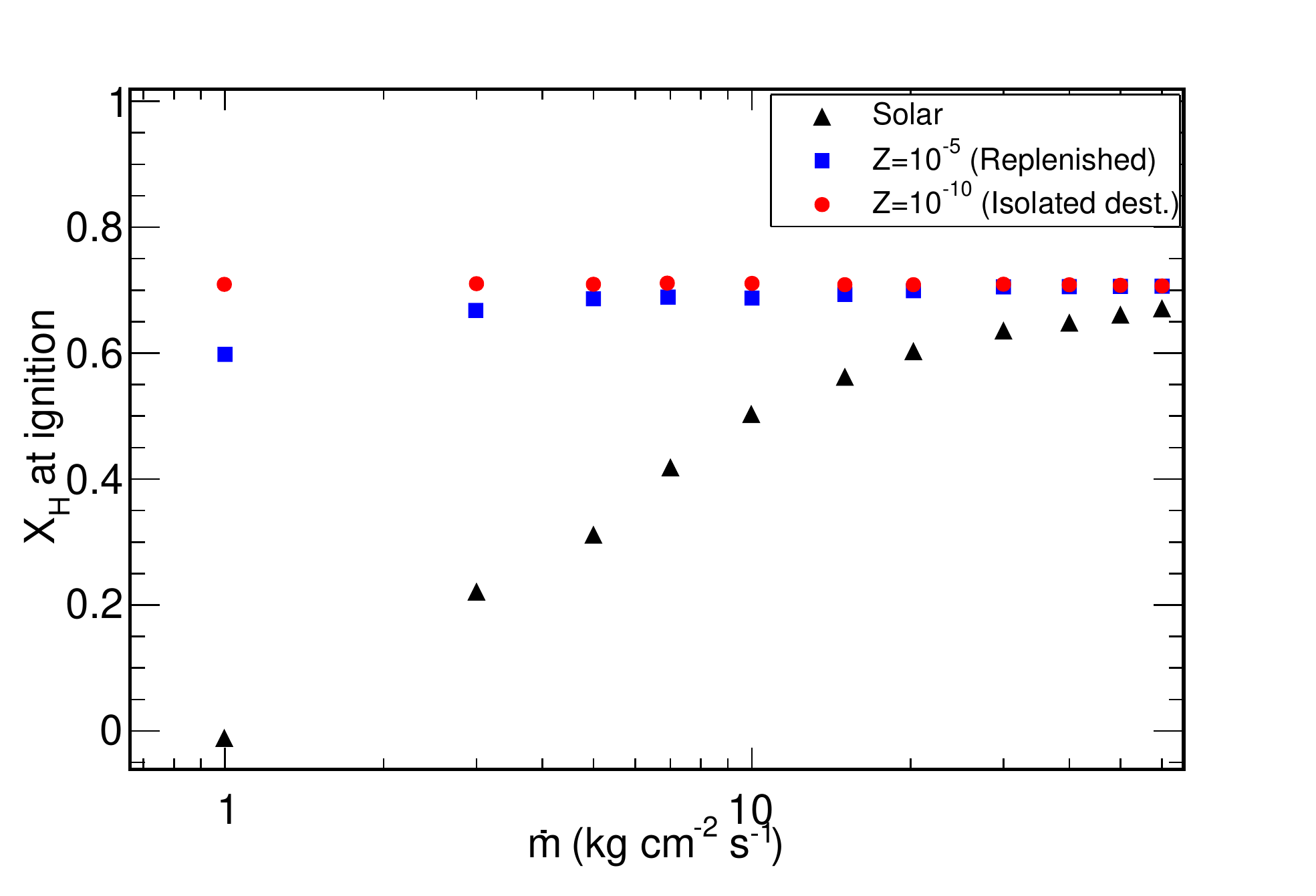}
\caption{Hydrogen mass fraction at the base of the accumulated column during ignition for different ${Z_{CNO}}$ values. We take M = 1.4 $\mathrm{M_{\odot}}$, R= 10 km and base flux F$_{b}=$0.15 MeV/$u$. These results are obtained using code \tt{settle}}
\label{figure4}
\end{center}
\end{figure}


\section{Discussion} \label{sec:cite}
\subsection{Impact on the burst ignition conditions}\label{}
The spallation calculations in the present work show that ${Z_{CNO}}$ can be as high as $10^{-5}$ (cascading model). To understand the impact of this ${Z_{CNO}}$ values on the X-ray burst ignition conditions, we have used the code {\tt settle}\footnote{https://github.com/andrewcumming/settle}, which computes ignition conditions for Type~I X-ray bursts using a multi-zone model of the accreting layer and a one-zone ignition criterion \citep{Cumming2000}. The hydrogen mass fractions present at the time of ignition as a function of mass accretion rate are shown in Figure~\ref{figure4} for different metallicities. Here, base flux has been assumed to be 0.15 MeV/$u$. Figure~\ref{figure4} shows that for solar $Z_{CNO}$, the amount of hydrogen present at the time of helium ignition decreases as the mass accretion rate decreases and after a certain point no hydrogen is left at the time of helium ignition leading to pure-helium bursts. For ${Z_{CNO}=10^{-10}}$ (isolated destruction), a negligible amount of hydrogen is burned before the burst and hydrogen is present during the helium ignition even at low accretion rates. Here, we find a substantial alteration of X(H) at low accretion rates between cascading/replenished and isolated destruction models. This significantly impacts the X-ray burst light curve and ash composition \citep{Meis19}. However, for $\dot{m} \sim$ 5 kg cm$^{-2}$ s$^{-1}$ and below, sedimentation affects the distribution of isotopes and the ignition of H and He in the envelope of an accreting neutron star \citep{Fang2007}. Further studies are required to see if this counteracts the effect of lower Z. Despite the differences, our results confirm the general conclusion of \citet{Bil92} that spallation leads to the burst ignition in a hydrogen-helium mixture for a broad range of accretion rates  even when a full cascading process is considered.      
\color{black}

\subsection{Impact of spallation-altered accreted composition on X-ray bursts ashes}

We performed multizone X-ray burst calculations with the code {\tt MESA}~\citep{Paxt11,Paxt13,Paxt15,Paxt18} to determine the impact of spallation in the atmosphere on the burst ashes. Calculations were performed following the methods of \citet{Meis18,Meis19}. 

Briefly, these consisted of discretizing an $0.01$~km atmosphere of a 1.4~$M_{\odot}$ 11.2~km neutron star into $\sim$1000~zones and following the nuclear burning and hydrodynamic evolution induced by accretion. The 304 isotope network of \citet{fisk08} and REACLIB~\citep{Cybu10} v2.2 reaction rates were used. Hydrodynamic corrections included a post-Newtonian modification of the local gravity to emulate general relativistic effects and convection was approximated using a time-dependent mixing length theory~\citep{Heny65,Paxt11}. MESA v9793 was used with the time resolution and spatial resolution adapting in time according to the {\tt MESA} controls {\tt varcontrol\_target=1d-3} and {\tt mesh\_delta\_coeff=1.0}~\citep{Paxt13}.

Accretion is achieved by adding a small amount of mass to the model's outer layers and readjusting the stellar structure~\citep{Paxt11}. An accretion rate of 10~kg$\,$cm$^{-2}$s$^{-1}$ was used along with a base heating of 0.15~MeV per accreted nucleon. Calculations were performed to investigate the impact of reduced metallicity and altered metal distribution.

\begin{figure}
\begin{center}
\includegraphics[width=\columnwidth,angle=0]{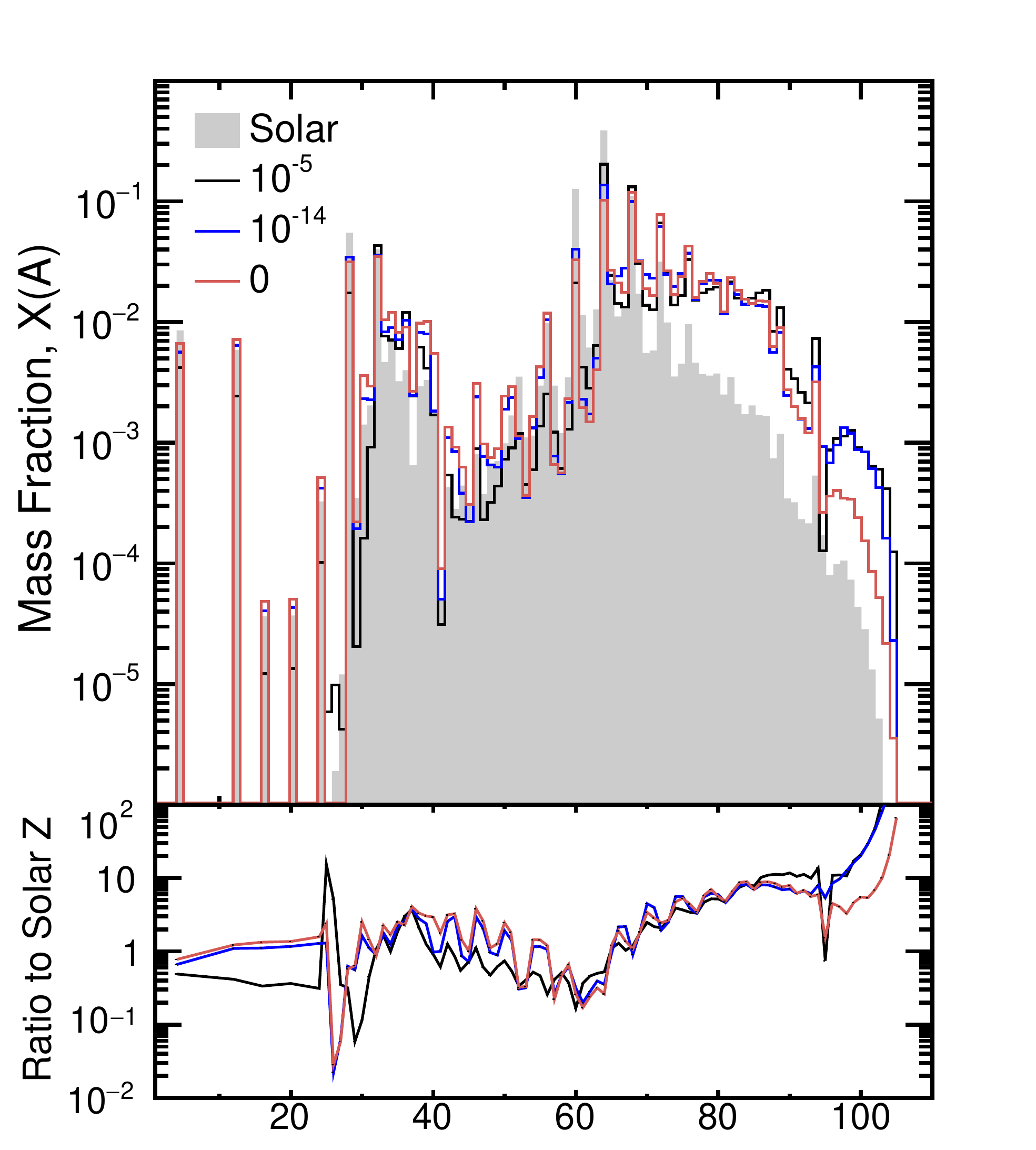}
\caption{(upper panel) Multi-zone calculation abundance distributions for $X=0.70$ and metallicity indicated by the legend. (lower panel) Ratio of X(A) results to calculations performed with solar $Z$.
\label{fig:MZashes}}
\end{center}
\end{figure}

For $X=0.70$ and $Y=1-X-Z$, calculations were performed with $Z=0.02$ (solar), $10^{-5}$ (cascading destruction/replenishment model), $10^{-14}$ (isolated destruction model), and $0$, each distributed over a solar metal distribution~\citep{Grev98}.
Results are shown in Figure~\ref{fig:MZashes}. For reduced $Z$, the abundance distribution is shifted to higher mass fractions. This is expected, as reduced CNO abundances reduce hydrogen burning during accretion, leading to more hydrogen rich conditions at burst ignition and therefore extended hydrogen burning~\citep{Heger07,Jose10,Meis19}. Between $Z=10^{-5}$ and $10^{-14}$, the mass fraction of isotopes summed by mass number (X(A)) can vary by more than a factor of 2 (Figure \ref{fig:MZashes}, lower panel). The lower $Z$ calculations show particularly differences in the oscillation pattern associated with the $A=4n$~\citep{Meis19} since the ignition conditions are even less helium-rich. These changes are significant relative to other changes in astrophysical conditions and nuclear reaction rates \citep{Meis19}. Most of the changes in X(A) are in the $A \sim 30-60$ region, which potentially affects the urca cooling neutrino luminosity in the crust \citep{Meis17}. In the inner crust of the neutron star the thermal conductivity is dominated by electron-ion impurity scattering. Ashes between $A= 29-55$ are funneled into $N=28$ shell closure, therefore changes in X(A) in this region (Figure \ref{fig:MZashes} (lower panel)) will potentially alter the impurity parameter~\citep{Lau18}.

\section{Conclusions}\label{cite:conclusion}
We provide the first calculations of the spallation-modified accreted composition for X-ray burst models using a full nuclear reaction network. The resulting composition is significantly different from the frequently used initial compositions based on the surface composition of the companion star. We find significant survival of CNO elements in the full cascading destruction model compared to previous isolated/non-replenished destruction estimates.  These larger CNO abundances, especially at lower accretion rates, alter the amount of hydrogen present at the time of burst ignition.We show that this has a significant effect on X-ray burst model calculations of the composition of the burst ashes. Our results are therefore important to take into account in X-ray burst model calculations that are used to predict the thermal and compositional structure of accreted neutron star crusts ~\citep{Meis19,Lau18}.

\acknowledgments
This work is mainly the product of discussions at the 3$^{rd}$ Astrophysical Reaction Network School supported by the National Science Foundation under Grant No. PHY-1430152 (JINA Center for the Evolution of the Elements). Z.M. was supported in part by the U.S. Department of Energy under grant Nos. DE-FG02-88ER40387 and DESC0019042. S.A.G. acknowledges support from the U.S. Department of Energy under Award Number DOE-DE-NA0002847 (NNSA, the Stewardship Science Academy Alliances program). H.S. acknowledge support from NSF  under PHY-1102511. B.S.M. was supported by NASA under grant No. NNX17AE32G.

%

\vspace{5mm}

\software{Nucnet, Settle \citep{Cumming2000},  
          \tt{MESA} \citep{Paxt11,Paxt13,Paxt15,Paxt18}
          }
\vspace{5mm}



\bibliography{references}{}
\bibliographystyle{aasjournal}



\end{document}